\documentstyle[manuscript,aps,epsf,array]{revtex}

\tightenlines

\newcommand{\beqa}{\begin{eqnarray}}
\newcommand{\eeqa}{\end{eqnarray}}
\newcommand{\beq}{\begin{equation}}
\newcommand{\eeq}{\end{equation}}

\newcommand{\bc}{\begin{center}}
\newcommand{\ec}{\end{center}}

\begin{document}

%
%
%
\draft
\title{
\vspace*{-2.cm}
\begin{flushright}
{\normalsize UTHEP-465}\\
{\normalsize UTCCP-P-131}\\
\end{flushright}
Analysis of unstable particle with the maximum entropy method
in $O(4)$ $\phi^4$ theory on the lattice
}

\author{\normalsize
  T.~Yamazaki$^{1}$ and
  N.~Ishizuka$^{1,2}$ \\
}

\address{
$^{\rm 1}$Institute of Physics,
    University of Tsukuba, Tsukuba, Ibaraki 305-8571, Japan \\
$^{\rm 2}$Center for Computational Physics,
    University of Tsukuba, Tsukuba, Ibaraki 305-8577, Japan 
}
\date{\today}

\maketitle

\begin{abstract}

We explore applications of the maximum entropy method (MEM) to 
determine properties of unstable particles using the four-dimensional 
$O(4)$ $\phi^4$ theory as a laboratory.  
The spectral function of the correlation function of the unstable $\sigma$ 
particle is calculated with MEM, and shown to yield reliable results 
for both the mass of $\sigma$ and the energy of the two-pion state. 
Calculations are also made for the case in which $\sigma$ is stable. 
Distinctive differences in the volume dependence of the $\sigma$ mass and 
two-pion energy for the stable and unstable cases are analyzed in terms of 
perturbation theory. 

\end{abstract}





In contrast to many successes in understanding of stable hadron states 
by numerical simulations,
unstable hadrons and 
hadronic decay processes are not understood well on the lattice.
An important example in QCD is $\rho\rightarrow\pi\pi$ decay.
There are a number of difficulties in dealing with such decay processes. 
One of the difficulties is computer power.
In full QCD lattice simulations accessible today,
the pion is so heavy that the $\rho$ meson cannot decay.
A more crucial problem is the difficulty first pointed out by Maiani and 
Testa~\cite{MT}.
Because the $\rho$ meson and the $\pi\pi$ state in the isospin $I=1$ channel
have the same quantum numbers,
the correlation function of the $\rho$ meson behaves 
as a multi-exponential function corresponding to $\pi\pi$ states with 
various relative momenta as well as the $\rho$ meson. 
In general, it is very difficult to decompose the states and extract the 
energy of each state from such a correlation function.

In this article we explore application of the maximum entropy method (MEM)
to extract the mass of unstable particle and the energies of states to which 
the unstable particle can decay. 
MEM has already been applied to extract the ground 
and first excited states of the stable mesons in QCD~\cite{hats:rev,MY}. 
The masses of the two states were simultaneously extracted from 
the peak positions of the spectral function of the meson.  
This experience leads us to expect MEM to be useful for 
the multi-exponential correlation function
of unstable particles such as the $\rho$ meson.

We employ the four-dimensional $O(4)$ $\phi^4$ theory for the present 
exploratory study. It is an effective theory of QCD, 
containing the pion and $\sigma$ particle. The $\sigma$ can be made 
either stable or unstable by choice of parameters. 
In the non-linear formulation with the 
constraint $\phi^{\alpha}(x)\phi^{\alpha}(x)=1$, the action is given by
\begin{equation}
S =
\sum_{x} \Bigl\{
- \kappa\cdot \phi^{\alpha}(x) D \phi^{\alpha}(x)
-J\phi^4(x)\Bigr\}
, 
\end{equation}
where $\phi^{\alpha}(x)=(\pi^i(x),\sigma(x))$ for $i=1,2,3$,
and
$D \phi(x) = \sum_\mu  \phi(x+\hat{\mu}) + \phi(x-\hat{\mu})$.
For $m_\sigma > 2 m_\pi$,
the $\sigma$ particle is unstable and
can decay to the $I=0$ two-pion state, similarly to the $\sigma$ meson in QCD.
We examine the efficiency of MEM for extracting the energies of $\sigma$ and 
$\pi\pi$ states for both the unstable $m_\sigma > 2 m_\pi$
and stable $m_\sigma < 2 m_\pi$  cases.
Very recently a similar study for the three-dimensional four-fermion model
has been carried out by Allton {\it et al.}~\cite{2+1}.

Numerical simulations are
carried out at $(\kappa,J)=(0.308,0.0012)$ and $(0.30415,0.003)$, 
corresponding to 
$m_{\sigma}/m_{\pi}\approx 3.7$
and $m_{\sigma}/m_{\pi}\approx 1.8$, 
for several spatial lattice sizes in the range $10^3-28^3$ 
to investigate the volume dependence of the spectral functions and energies.
The temporal lattice size is fixed to $T=64$.
Configurations are generated by 
the multi-cluster algorithm~\cite{O4}.
The periodic boundary condition is imposed in all directions.
We perform $0.6 \times 10^6$ -- $1.2\times 10^6$ iterations per simulation
point ($\kappa,\ J$) and volume.

We calculate the correlation function matrix~\cite{O4,ph1}
given by 
$C_{ij}(\tau)=\langle (O_i(\tau)-O_i(\tau+1))O_j(0)\rangle$
for $i,j=\sigma$ and $\pi\pi$, where
$O_{i}$ is the interpolating operator either 
for the $\sigma$ or the $I=0$ two-pion state with zero momentum,
and the subtraction $O_i(\tau)-O_i(\tau+1)$ is made 
for eliminating the vacuum contribution.
We apply MEM to the diagonal parts of the correlation function matrix
$C_{ii}(\tau)$.

\begin{figure}[h]
\leavevmode
\centerline{\epsfxsize=6cm\epsfbox{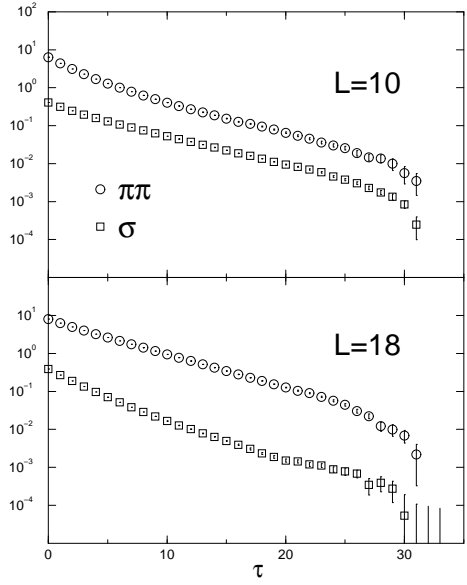} }
\caption{
Correlation functions for the $\sigma$ and $\pi\pi$
in the unstable case.
\label{fig:corre}
}
\end{figure}

In Fig.~\ref{fig:corre} we illustrate the correlation
functions in the $\sigma$ and $\pi\pi$ channels for two volumes 
in the unstable case.
The errors are estimated by the jackknife method
with the bin size of 8000 to 16000 iterations.
For the smaller volume the $\pi\pi$ correlation function exhibits a change of 
slope at $t\approx 10$, from a larger slope corresponding to $\sigma$ to a 
smaller one of the $\pi\pi$ state. 
The $\sigma$ correlator decays parallel to the latter, 
indicating dominance of the $\pi\pi$ state for small volumes. 
For the larger volume, the trend is opposite, the $\pi\pi$ correlation 
function showing a change of slope from the $\sigma$ particle to 
$\pi\pi$ state at $t\approx 20$, while the $\pi\pi$ correlator is dominated 
by the $\pi\pi$ state. 
The decrease of ``off-diagonal'' contributions with increasing volume 
in the correlation functions can be understood
from a perturbative analysis: the overlaps
$|\langle 0|\sigma|\pi\pi\rangle|^2$ and $|\langle 0|\pi\pi|\sigma\rangle|^2$
are proportional to
$
{1}/{[ L^3(m_{\sigma}-E_{\pi\pi})^2 ]}
$
where $E_{\pi\pi}$ is the two-pion energy.

The spectral function $f(\omega)$ in the $\sigma$ and $\pi\pi$ channels 
is defined in terms of the correlation function $C_{ii}(\tau)$ 
($i=\sigma,\pi\pi$) through
\begin{eqnarray}
L^3C_{ii}(\tau)&=&
\int\! d\omega f_{i}(\omega)K(\omega,\tau)
,
\end{eqnarray}
where
$K(\omega,\tau)=
e^{-\tau \omega}(1-e^{-\omega})+e^{\omega(\tau-T)}(1-e^{\omega})$.
In our MEM analysis the model function $m(\omega)$ is chosen as
$m(\omega)=A(\omega_0) \cdot \omega_0^4 / \omega^4$, 
where $\omega_0$ is a reference point and 
$A(\omega_0)$ is the perturbative $\sigma$ spectral function
at a reference point $\omega_0$, which is given by 
\beq
A(\omega_0) = 
\frac{M_I (\omega_0)}
{\pi\bigl[\bigl(
\omega_0^2-\widetilde{m}_{\sigma}^2+M_{\! R}(\omega_0)\bigr)^2
+M^2_I (\omega_0)\bigr]} / Z_{\sigma}.
\label{eq:A}
\eeq
The functions $M_{\! R}(\omega_0)$ and $M_I(\omega_0)$
are defined by
\begin{eqnarray}
&& 
M_{\! R} (\omega_0) =C[
\gamma_{\pi}(\omega_0)-\gamma_{\pi}(\widetilde{m}_{\sigma})
+3\gamma_{\sigma}(\omega_0)],
\label{eq:MR}
\\
&& 
M_I (\omega_0) =
C\pi[\beta_{\pi}(\omega_0) +3\beta_{\sigma}(\omega_0)],
\label{eq:MI}
\end{eqnarray}
where
$\gamma_{\alpha}(\omega_0)=\beta_{\alpha}(\omega_0)
\ln[
(1+\beta_{\alpha}(\omega_0))/(1-\beta_{\alpha}(\omega_0)]$
and
$
\beta_{\alpha}(\omega_0) = \sqrt{1-(4\widetilde{m}_{\alpha}^2/
\omega_0^2)}\,\theta(\omega_0-2\widetilde{m}_{\alpha})
$
for $\alpha=\sigma$ and $\pi$.
The constant $C$ in Eqs.~(\ref{eq:MR}) and (\ref{eq:MI}) 
is given
by $C=3Z_{\pi}(\widetilde{m}^2_{\sigma}-\widetilde{m}^2_{\pi})^2/32\pi^2v^2$,
where $v=\langle \sigma\rangle$,
and $Z_{\alpha}$ ($\alpha=\sigma,\pi$) 
is the wave function renormalization factor 
for $\sigma$ and $\pi$, respectively.
We apply the same model function to the $\sigma$ and 
$\pi\pi$ correlation functions.

We choose $\omega_0=2$, 
and $\widetilde{m}_{\alpha}$ and $Z_{\alpha}$ in Eq.(\ref{eq:A})
are fixed to the values estimated from 
inverse correlation functions in momentum space.
The input parameters for the model function 
are compiled in Table~\ref{tab:param}.
The reconstruction is carried out in the region $0\leq \omega\leq 3$.
We choose $\Delta\omega=5\times 10^{-4}$ around the peaks corresponding to 
the $\sigma$ and $\pi\pi$ states to determine the energies accurately,
and $\Delta\omega=10^{-2}$ in other regions.
The number of data is taken as large as possible $0\le \tau \le T/2$.
We also check that the final results for the spectral functions 
do not depend much on the model function.

\begin{figure}[h]
\centerline{\epsfxsize=7cm \epsfbox{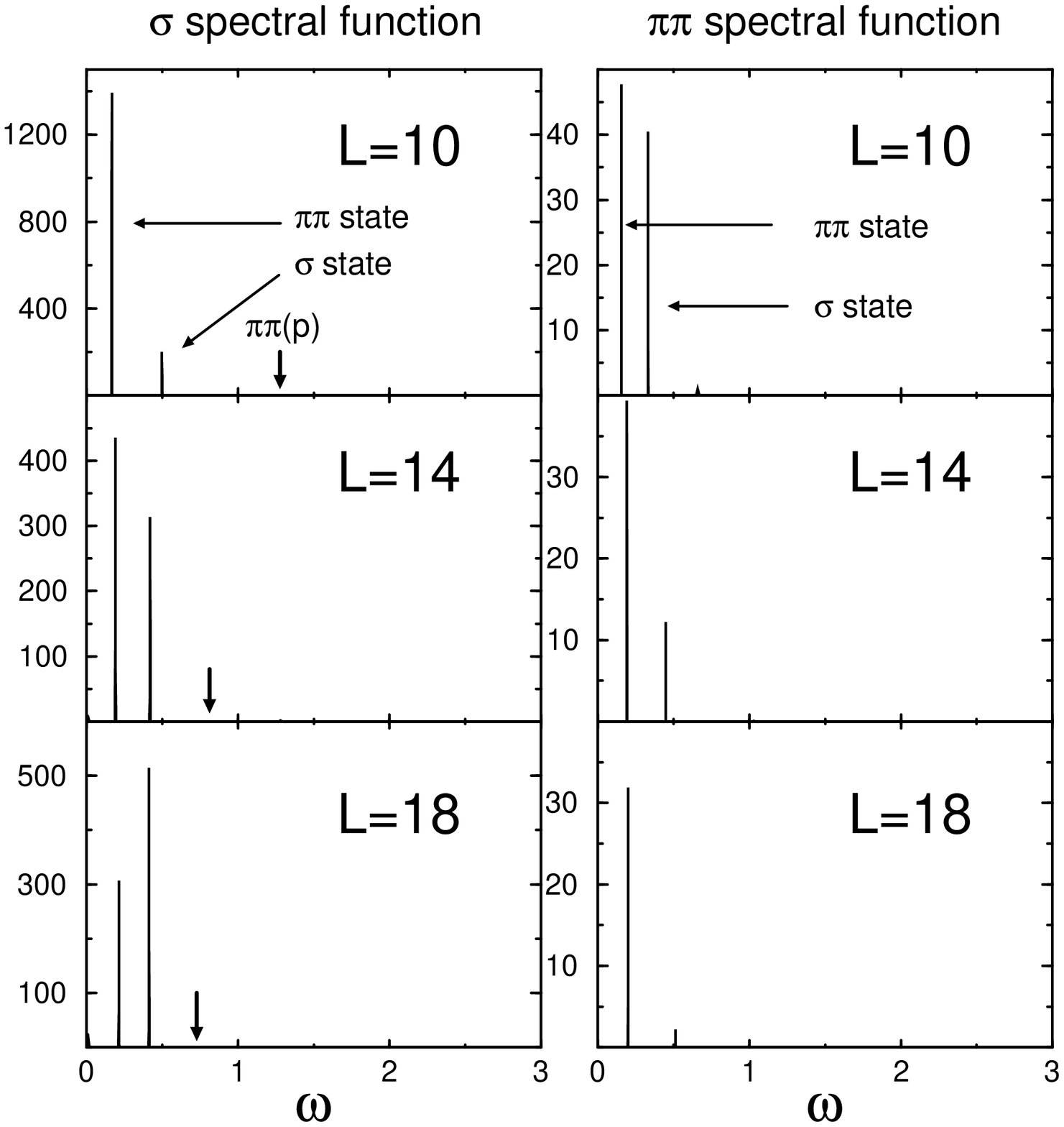} }
\caption{
Spectral functions reconstructed from 
$\sigma$ (left line) and $\pi\pi$ (right line)
correlation functions
in the unstable case.
\label{fig:spe_unstable}
}
\end{figure}

In Fig.~\ref{fig:spe_unstable} we show 
the spectral functions for the $\sigma$ and $\pi\pi$ correlation functions, 
$f_\sigma(\omega)$ and $f_{\pi\pi}(\omega)$,
in the unstable case. 
Since energy eigenvalues are discrete on a finite volume,
the spectral function is a sum of $\delta$ functions.
We indeed observe sharp peaks for both spectral functions,
which can be identified as the $\pi\pi$ state with zero momentum at 
the first peak and the $\sigma$ state at the second peak.
The decrease of the peak height for the $\pi\pi$ state in $f_\sigma(\omega)$ 
and for the $\sigma$ state in $f_{\pi\pi}(\omega)$ agrees with 
the volume dependence of the correlation functions discussed above. 
In the figure for $f_\sigma(\omega)$
the position of peak expected for the $\pi\pi$ state with momentum $p=2\pi/L$ 
is indicated by a downward arrow.
The absence of peak in our data implies that 
the overlap of $\sigma$ with this state is very small.

\begin{figure}[h]
\centerline{\epsfxsize=7cm \epsfbox{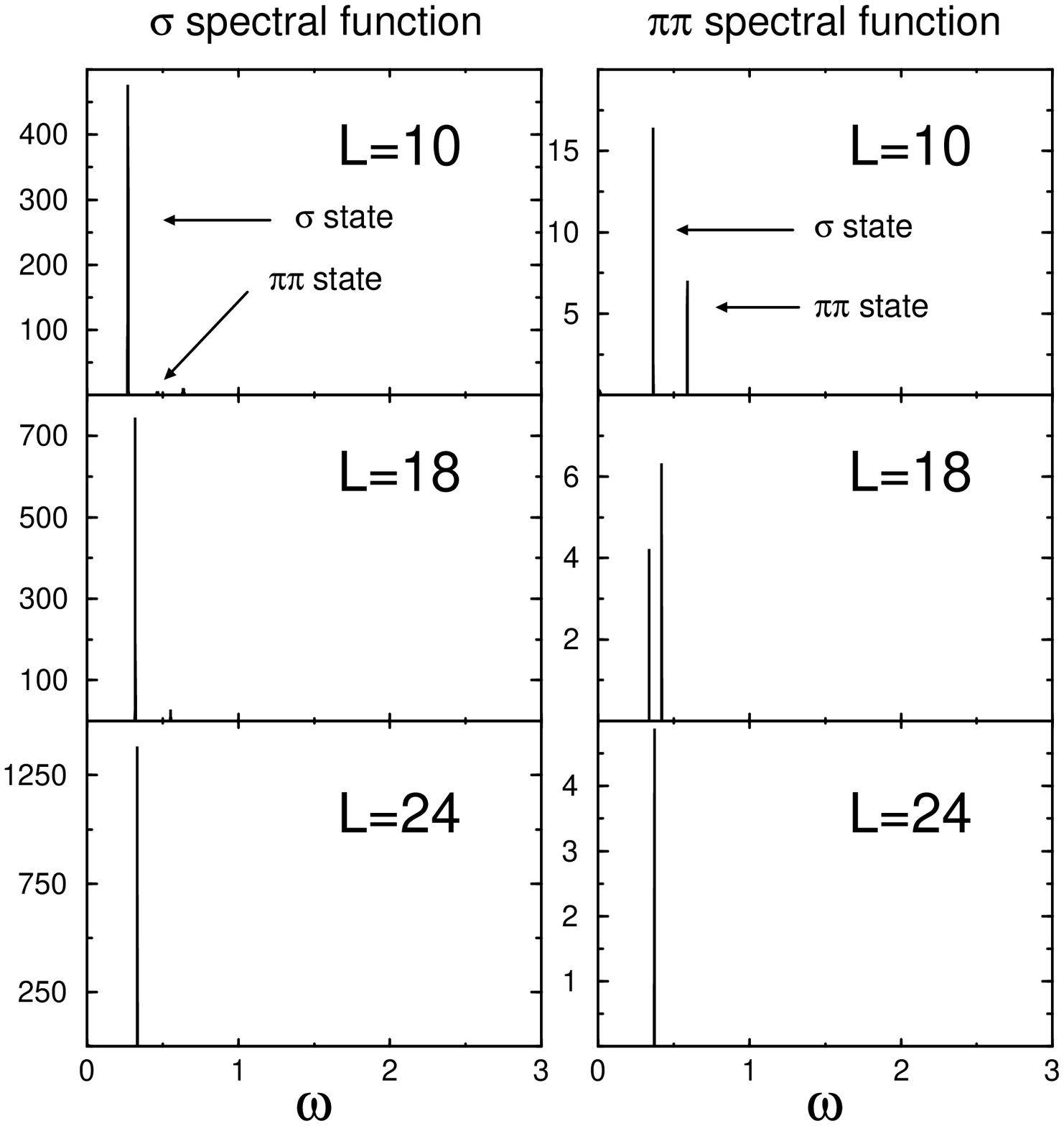} }
\caption{
Spectral functions reconstructed from 
$\sigma$ (left line) and $\pi\pi$ (right line)
correlation functions
in the stable case.
\label{fig:spe_stable}
}
\end{figure}

Fig.~\ref{fig:spe_stable} shows the spectral functions in the stable case. 
For $f_{\pi\pi}(\omega)$ the $\sigma$ contribution decreases with volume 
similar to the unstable case. 
We observe only a single peak in $f_\sigma(\omega)$, 
indicating that $|\langle 0|\sigma|\pi\pi\rangle|^2$ is very small
in this case.

\begin{figure}[h]
\centerline{\epsfxsize=6.5cm \epsfbox{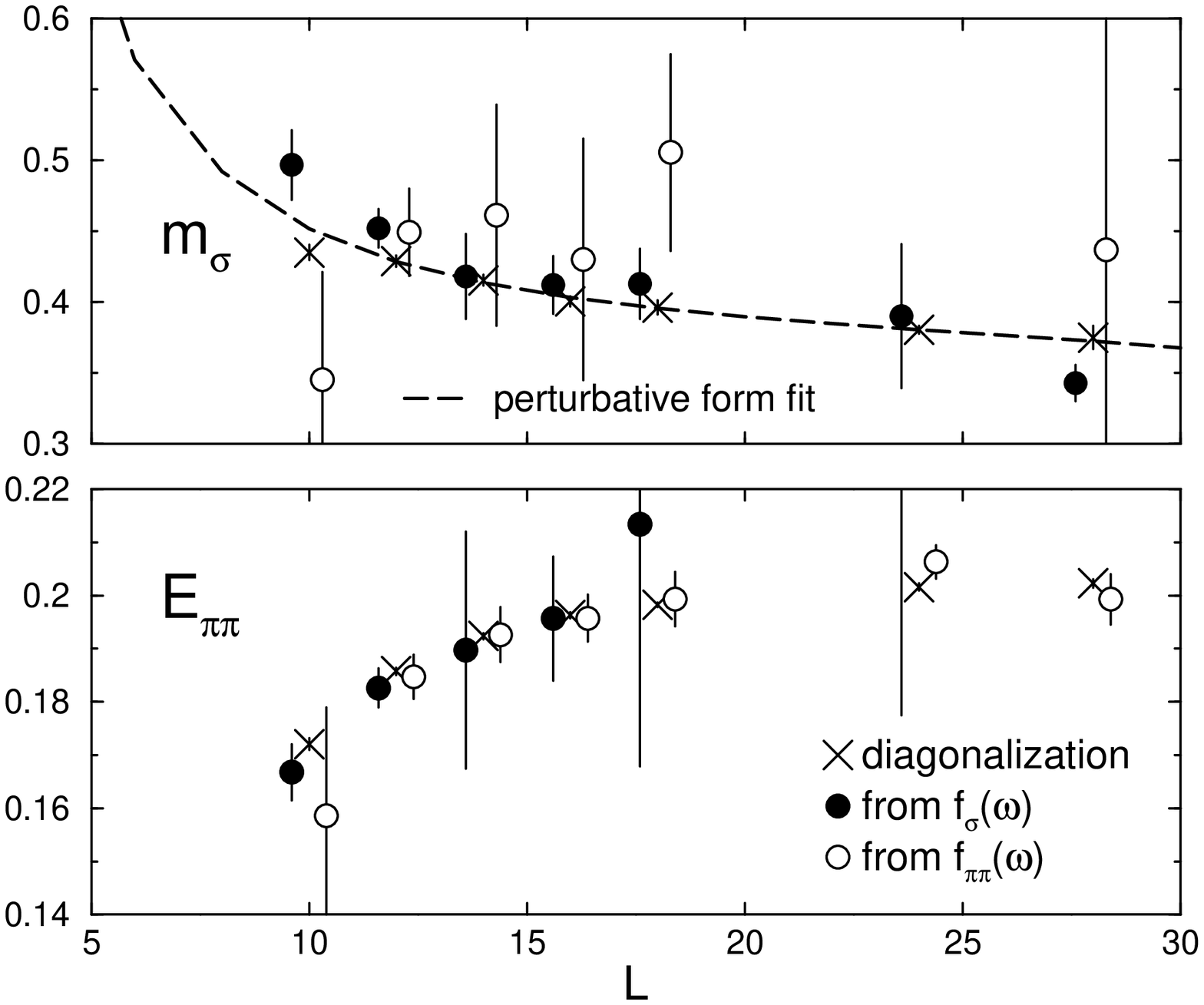} }
\caption{
Energies for $\sigma$ and $\pi\pi$ with the MEM and
diagonalization in the unstable case.
\label{fig:energy_decay}
}
\end{figure}

\begin{figure}[h]
\centerline{ \epsfxsize=6.5cm \epsfbox{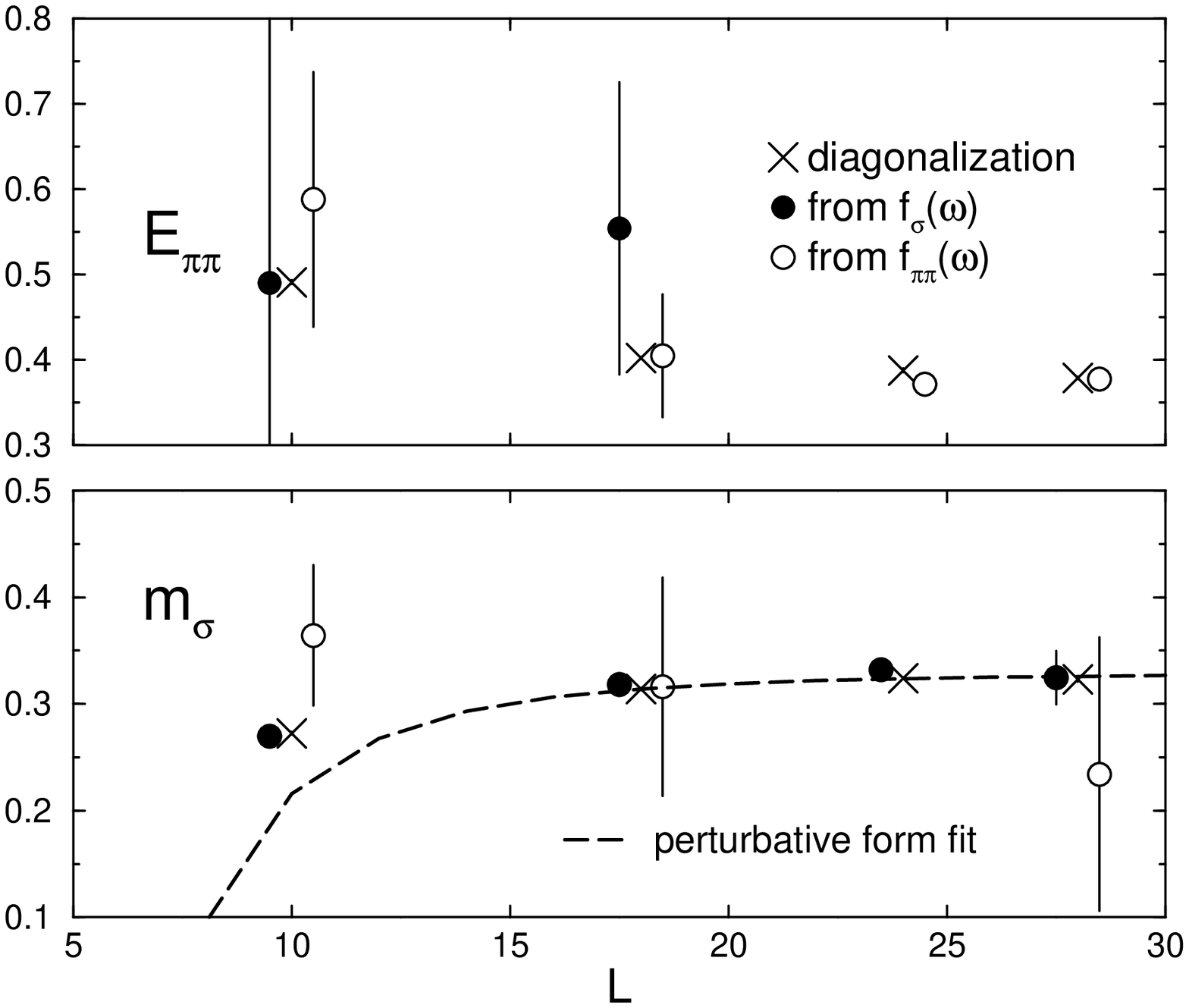} }
\caption{
Energies for $\sigma$ and $\pi\pi$ with the MEM and
diagonalization in the stable case.
\label{fig:ene_scat}
}
\end{figure}

The $\sigma$ mass $m_{\sigma}$ and the $\pi\pi$ state energy
$E_{\pi\pi}$ obtained from the peak positions of the 
spectral functions as functions of spatial lattice size $L$ are shown 
in Fig.~\ref{fig:energy_decay}(unstable case) and \ref{fig:ene_scat} 
(stable case) by circles. 
For larger volumes 
$m_{\sigma}$ from $f_{\pi\pi}(\omega)$ and 
$E_{\pi\pi}$ from $f_{\sigma}(\omega)$ 
suffer from large errors or are not available. 
This is because the overlaps 
$|\langle 0|\sigma|\pi\pi\rangle|^2$ and 
$|\langle 0|\pi\pi|\sigma\rangle|^2$
decrease as the volume increases.

We have seen in Figs.~\ref{fig:spe_unstable} and \ref{fig:spe_stable}
that the $\sigma$ and $\pi\pi$ correlation functions are 
dominated by the $\sigma$ and $\pi\pi$ states with zero momentum, 
and other states are negligible.
In this case we can apply the diagonalization method~\cite{ph1} to the 
2$\times$2 correlation function matrix $C(\tau)$
for extraction of the energy 
eigenvalues of these states.
We diagonalize the matrix
$D(\tau,\tau_0)=C^{-1/2}(\tau_0)\ C(\tau)\ C^{-1/2}(\tau_0)$ 
at each $\tau$, where $\tau_0$ is some reference time chosen to be 
$\tau_0=0$ in this work.
The eigenvalue of $D(\tau,\tau_0)$
is given by $\lambda_{\nu}(\tau,\tau_0)=K(W_{\nu},\tau)/K(W_{\nu},\tau_0)$
with $K(W_{\nu},\tau)=e^{-\tau W_{\nu}}(1-e^{-W_{\nu}})
+e^{W_{\nu}(\tau-T)}(1-e^{W_{\nu}})$,
where $W_{\nu}$ is the energy of the states $\nu=\sigma,\pi\pi$.

The energies obtained by the diagonalization method 
are plotted by cross symbols 
in Figs.~\ref{fig:energy_decay} and \ref{fig:ene_scat}.
The errors are estimated by the
jackknife method with 8000 to 16000 eliminations. 
The results for $m_\sigma$ and $E_{\pi\pi}$ 
are consistent with those with MEM, but   
the statistical error with the diagonalization
method is much smaller.
This is because the diagonalization makes full use of the $2\times 2$ 
correlation function matrix while MEM utilizes on the diagonal element. 
 
For the application of the diagonalization method, 
it is essential that the dominant states in the correlation functions are
known. Otherwise we need to consider
the correlation function matrix of all possible states 
with the same quantum numbers, which is a difficult task.
MEM is useful for identifying the dominant states, as exemplified with 
our example. 

Comparing Fig.~\ref{fig:energy_decay} with Fig.~\ref{fig:ene_scat}
we find an essential difference in the volume
dependence of the $\sigma$ mass and $\pi\pi$ energy
between the unstable and stable cases.
In the unstable case,
the $\pi\pi$ energy increases and 
the $\sigma$ mass decreases as the volume increasing, 
while an opposite trend is seen in the stable case.

The volume dependence of the $\sigma$ mass 
expected from perturbation theory is given by 
\beqa
m_{\sigma}(L,m_{\sigma},g_R) &=& m_{\sigma} + g_R
(\Delta m_{\sigma}(L)-\Delta m_{\sigma}(\infty)).
\label{eq:m_sigma_L}
\eeqa
Here $m_{\sigma}$ is the $\sigma$ mass for infinite volume,
$g_R$ is the renormalized coupling constant, and 
\beqa
\Delta m_{\sigma}(L) &=&
\frac{1}{4m_{\sigma}L^3}
\sum_{\vec{p}}\sum_{\alpha=\pi,\sigma}
\left[
\frac{D}{W_{\alpha}(p)}
+A_{\alpha} C_{\alpha}(p)
\right],\label{eq:voldep}
\eeqa
with
$A_{\alpha}=2,6$ for $\alpha=\pi,\sigma$, 
$D=
1-{3(m_{\sigma}^2-m_{\pi}^2)}/{m_{\sigma}^2}
$, 
and 
\begin{equation}
C_{\alpha}(p) =
\frac{m_{\sigma}^2-m_{\pi}^2}
{W_{\alpha}(p)(m_{\sigma}^2-W_{\alpha}^2(p))},
\end{equation}
where 
$W_{\alpha}(p) = 2\sqrt{m_{\alpha}^2 + 4 \sum_{i=1}^3 \sin^2(p_i/2)}$
with $p_i$ being the spatial momenta. 

We fit our results for the $\sigma$ mass 
obtained by the diagonalization method
to Eq.(\ref{eq:m_sigma_L}), taking $m_{\sigma}$ and $g_R$ as 
fit parameters and setting $m_{\pi}$ to the value 
obtained from the pion correlation function at $L=28$.
The fit curves are plotted 
in Fig.~\ref{fig:energy_decay} and Fig.~\ref{fig:ene_scat}
for each case,
where the data at $L=10$ are excluded from the fitting.
The fit parameters and $\chi^2$ are compiled in Table~\ref{tab:volfit}.
The fits agrees quite well with the simulation results.
We then realize that the different volume dependence 
between the unstable and stable cases
originates from an opposite sign of $C_{\pi}(p)$ in Eq.(\ref{eq:voldep})
in the two cases.

In order to understand the volume dependence of the $\pi\pi$ energy,
we consider the scattering length $a_0$. 
It is related to the energy shift of the two-pion state 
through L\"uscher's formula~\cite{Lu1} given by 
\begin{equation}
E_{\pi\pi}-2m_{\pi}
= -\frac{4\pi a_0}{m_{\pi}L^3}
\left(1+c_1\frac{a_0}{L}+c_2\frac{a_0^2}{L^2}\right),  
\end{equation}
where
$c_1 = -2.837297,\ c_2 = 6.375183$.
The results for $a_0$
obtained from $E_{\pi\pi}$ calculated with the diagonalization method
are tabulated in Table~\ref{tab:scat-pert} in the column entitled 
``Simu. diago.''.
Here $m_{\pi}$ is fixed to the value at $L=28$.
The sign of $a_0$ differ in the two cases, 
which reflects the difference in the volume dependence of $E_{\pi\pi}$.

We can estimate the scattering length 
in perturbation theory.  From results for the perturbative phase 
shift~\cite{O4} we obtain
\beq
a_0 = 
\frac{g_R m_{\pi}}{96\pi m_{\sigma}^2}
\Bigl(7R^2+8\Bigr)
\frac{1}{R^2-4},
\label{eq:scatlength}
\eeq
where $R = m_{\sigma}/m_{\pi}$.
To evaluate the right handside, we use the fit result for 
$m_{\sigma}$ given in Table~\ref{tab:volfit}.
The coupling constant $g_R$ is obtained in two ways, 
either from the perturbative definition~\cite{O4,Lu0} denoted as 
$g_R({\rm def.})=3Z_{\pi}(m^2_{\sigma}-m^2_{\pi})/v^2$ or 
from the fit of the volume dependence of the $\sigma$ mass given in
Table~\ref{tab:volfit} denoted as $g_R$(fit).
For $m_{\pi}$, $Z_{\pi}$ and $v$ we use 
the results at $L=28$ as before.

In Table~\ref{tab:scat-pert} the simulation results for $a_0$ are 
compared with the two estimates using the perturbative formula 
Eq.~(\ref{eq:scatlength}). 
While we cannot claim a precise agreement, we observe consistency 
in the value and sign obtained with simulation and perturbation theory
for both the unstable and stable cases.
Thus it is the factor of $R^2-4$ in the perturbative formula 
which leads to the the opposite volume dependence of the $\pi\pi$ energy 
in the two cases found by the simulation.

In this paper,
we have investigated the efficiency of 
the maximum entropy method 
for study of unstable particle systems using 
the four-dimensional $O(4)$ $\phi^4$ theory.
We have demonstrated that
the $\sigma$ mass and $\pi\pi$ energy 
can be obtained from the $\sigma$ correlation function alone.
We have also explained that 
the difference in  
the volume dependences of the
$\sigma$ mass and $\pi\pi$ energy
between the unstable 
and stable cases can be understood 
by perturbation theory.
It is an advantage of MEM that
only the single particle correlation function of the unstable particle is 
needed to analyze both the particle itself and the multi-particle decaying 
states.  
Furthermore it works even when the dominant states in the correlation 
functions are not known.
We expect the MEM analysis to play a useful role in 
future studies of unstable particles and decays in lattice QCD.
\hfill\break

This work is supported in part by Grants-in-Aid of the Ministry of Education 
No.~12740133.

\begin{table}[h]
\begin{center}
\begin{tabular}{llllll}\hline\hline
\multicolumn{6}{l}{unstable case}\\ \hline
L&$\widetilde{m}_{\pi}$&Z$_{\pi}$&$\widetilde{m}_{\sigma}$&Z$_{\sigma}$
&$v$\\\hline
10 &0.1138(2)&0.961(1) &0.309(2) &0.956(1) &0.112921(7)\\ 
12 &0.1084(2)&0.9657(8)&0.337(2) &0.949(1) &0.127042(4)\\
14 &0.1053(2)&0.9667(6)&0.351(1) &0.944(1) &0.13351169(4)\\
16 &0.1041(3)&0.9674(6)&0.359(1) &0.938(1) &0.136818(1) \\
18 &0.1036(2)&0.9676(6)&0.362(1) &0.935(1) &0.138598(1)\\
24 &0.1024(2)&0.9678(8)&0.365(1) &0.928(1) &0.140646(2)\\
28 &0.1024(2)&0.9685(7)&0.365(1) &0.925(2) &0.1410983(6)\\\hline
\end{tabular}
\begin{tabular}{llllll}\hline
\multicolumn{6}{l}{stable case}\\ \hline
L&$\widetilde{m}_{\pi}$&Z$_{\pi}$&$\widetilde{m}_{\sigma}$&Z$_{\sigma}$
&$v$\\\hline
10 &0.1945(2)&0.9734(4)&0.328(1) &0.967(1) &0.099507(1) \\
18 &0.1851(2)&0.9754(4)&0.3320(7)&0.951(1) &0.1094956(8)\\
24 &0.1844(2)&0.9735(5)&0.3344(6)&0.949(1) &0.1099262(1)\\
28 &0.1848(2)&0.9755(5)&0.3324(7)&0.945(1) &0.110067(2)\\ \hline\hline
\end{tabular}
\end{center}
\caption{
The input parameters for the model function of the MEM
in the unstable and stable cases.
\label{tab:param}
}
\end{table}

\begin{table}[h]
\begin{center}
\begin{tabular}{ccc}\hline\hline 
            &unstable              &stable   \\ \hline
$m_{\sigma}$&0.3765(2)             &0.3285(1)\\ 
$g_R$       &14(1)                 &9(1)\\ 
$\chi^2$/d.o.f.&0.23               &2.5\\ \hline\hline
\end{tabular}
\end{center}
\caption{
Fit parameters and $\chi^2$/d.o.f. (degrees of freedom)
of the volume dependence of the $\sigma$ mass.
\label{tab:volfit}}
\end{table}

\begin{table}[h]
\begin{center}
\begin{tabular}{lcccc}\hline\hline
&\multicolumn{2}{c}{unstable}&
\multicolumn{2}{c}{stable}\\ \hline
&$a_{0}$&$g_R$&$a_{0}$&$g_R$\\ \hline
Simu. diago.&
0.289(9) &       &$-2.49(19)$ &\\ 
Pert. $g_R({\mathrm def.})$&
0.494(1) &19.1(2)&$-3.62(11)$ &17.8(1)\\
Pert. $g_R$(fit)&
0.361(3) &14(1)  &$-1.98(23)$ &9(1)   \\ \hline\hline
\end{tabular}
\end{center}
\caption{
Scattering lengths
for the the simulation (Simu.) and perturbative (Pert.) results.
\label{tab:scat-pert}
}
\end{table}

\end{document}